\documentstyle[aps,amstex,amssymb,eqsecnum,epsf,psfrag]{revtex}

\begin{document}

\title{Some mathematical problems in numerical relativity}

\author{Maria Babiuc${}^{1}$ ,
	B\'{e}la Szil\'{a}gyi${}^{1}$ ,
	Jeffrey Winicour${}^{1,2}$
       }
\address{
${}^{1}$ Department of Physics and Astronomy \\
         University of Pittsburgh, Pittsburgh, PA 15260, USA \\
${}^{2}$ Max-Planck-Institut f\" ur
         Gravitationsphysik, Albert-Einstein-Institut, \\
	 14476 Golm, Germany
}
	 
\maketitle

\section{Introduction}

The main goal of numerical relativity is the long
time simulation of highly nonlinear spacetimes that
cannot be treated by perturbation theory. There are
three elements to achieving this:

(1) Analytic issues, such as well-posedness,
constraints, boundary conditions, linear stability,
gauge conditions and singularity avoidance.

(2) Computational issues, such as evolution and
boundary algorithms, numerical stability,
consistency, spacetime discretization and numerical
dissipation.

(3) Physical issues, such as simulation of the
desired global spacetime, extraction of the
radiation from an isolated system, the proper
choice of initial data, long timescale evolutions
tracking many orbits of an inspiraling binary.  

The correct treatment of the physical issues introduces severe global
problems. For instance, long term simulations are needed to flush out the
spurious gravitational radiation contained in the initial data for the
gravitational field of a binary black hole so that a physically relevant
waveform can be extracted. Furthermore, extraction of the waveform requires
a compactified grid extending to null infinity, or some alternative
approximation based upon an outer boundary in the radiation zone.

At present, the major impasses to achieving such global simulations are of
an analytical/computational nature. We present here some examples of how
analytic insight can lend useful guidance for the improvement of numerical
approaches.

\section{Waves}
\label{sec:waves}

The prime physical objective is to compute the gravitational waves emanating
from a compact source. We begin by introducing some underlying mathematical
and computational problems in terms of two examples of scalar wave
propagation. Both of these examples are chosen because they have a direct
analogue in general relativity and illustrate computational problems that
arise because of exponentially growing modes in an analytic problem
which is well-posed.

\subsection{Unbounded exponential growth}

First, consider a nonlinear wave propagating in Minkowski space according to
\begin{equation}
     \eta^{\alpha\beta}\partial_\alpha \partial_\beta \Phi
            -\frac{1}{\Phi} \eta^{\alpha\beta} (\partial_\alpha \Phi )
               (\partial_\beta \Phi)=0
      =\Phi\eta^{\alpha\beta}\partial_\alpha \partial_\beta \log \Phi.
\label{eq:logwave}
\end{equation} Although this nonlinear equation arises from a linear wave
equation for $\log \Phi$, it is a remarkably accurate model for some of the
problems that occur in numerical relativity. In order to simplify the problem,
we first impose periodic boundary conditions so that the evolution takes place
in a 3-torus $T^3$, i.e. on a boundary free manifold. For $\Phi>0$ the Cauchy
problem for this system is well-posed. Furthermore, the linear superposition
$\log \Phi_1 + \log \Phi_2$ of solutions to the linear wave equation correspond
to the solution $\Phi_1\Phi_2$ of the nonlinear problem.

A nonsingular solution of this system is the wave
\begin{equation}
         \Phi=1+F(t-z),
\label{eq:travwave}	 
\end{equation} 
where $F>-1$. Suppose we try to simulate this solution numerically. If
numerical error excites an exponentially growing mode of this system then
noise in this mode will eventually dominate the wave being simulated. For the
linear wave equation there are no such exponential modes. But this
nonlinear system admits the solutions
\begin{equation}
         \Phi_\lambda=  e^{\lambda t} (1+F(t-z)),
	 \label{eq:expt}
\end{equation}
for arbitrary $\lambda$.

Thus, although we have a well-posed initial value problem whose principle
part is the Minkowski wave operator, the simulation of a simple traveling
wave is complicated by the existence of neighboring solutions which grow
exponentially in time. Numerical error will excite these exponentially
growing modes and eventually dominate the traveling wave we are attempting
to simulate. You might ask: Why not choose $\log \Phi$ as the evolution
variable? This would clearly solve all numerical problems. As we have
already said, we have chosen this example because it arises in numerical
relativity where there is no analogous way to take the logarithm of the
metric. But there are indirect ways to model the derivative of a logarithm,
analogous to grouping terms according to $\Phi^{-1}\partial \Phi$
and rewriting the nonlinear wave equation (\ref{eq:logwave}) as
\begin{equation}
    \eta^{\alpha\beta}\partial_\alpha(\frac{1}{\Phi}\partial_\beta \Phi )=0.
\end{equation} This indeed works for the scalar field, as illustrated in Fig.
\ref{fig:scalar}. (See the discussion of finite difference methods in
Sec.~\ref{sec:gr}.) We will come back to the gravitational version of this
problem but first we give another example which illustrates a similar
complication with the {\em initial-boundary} value problem. 

We base this example on the initial-boundary value problem (IBVP) for this
nonlinear scalar field in the region $-L\le z\le L$ obtained by opening up
the 3-torus to $T^2\times R$. We consider the simulation of a traveling wave
wave packet $\Phi=1+F(t-z)>0$ with the Neumann boundary condition
$\partial_z \Phi|_{z=\pm L} =\partial_z F|_{z=\pm L}$.  The wave packet gets
the correct Neumann boundary data for it to enter the boundary at $z=-L$,
propagate across the grid and exit the boundary at $z=+L$.

In the process, numerical noise will be generated. There are
solutions of the system of the form
\begin{equation}
         \Phi_\epsilon=  \epsilon^2e^{t/\epsilon} \bigg(1+f(t-z)\bigg),
	 \label{eq:expt2}
\end{equation}
for arbitrary $\epsilon>0$.  Normally, if a scalar field admitted such
solutions we could infer that the corresponding Cauchy problem was ill-posed by
arguing (following Hadamard) that the solution $\Phi_0 =0$ has vanishing Cauchy
data and that the neighboring solutions $\Phi_\epsilon$ have unbounded size for
any $t>0$. However, the above Cauchy problem is well-posed because the initial
data must satisfy $\Phi>0$.

After the wave packet has crossed the grid, the remnant numerical noise gets
homogeneous Neumann data $\partial_z \Phi|_{z=\pm L} =0$. Thus it is reflected
off the boundaries and trapped in the simulation domain where it can grow
exponentially. The noise is generated while the wave packet is traveling across
the grid. The short wavelength modes can be controlled by introducing numerical
dissipation. But the long wavelength modes cannot be damped without the danger
of damping the signal. Just as in the case of periodic boundary conditions,
numerical error can excite exponential modes that destroy the accuracy of
a simulation in the case of Neumann boundary conditions.

You might ask: Why use Neumann boundary conditions? The Sommerfeld boundary
condition $(\partial_t\pm  \partial_z)\Phi |_{z=\pm L} =0$ does not admit
such modes and moreover it propagates numerical noise off the grid. The
answer to that question has to wait until  we have discussed the constraint
equations of general relativity.

The problem with using Neumann boundary conditions is not of analytic
origin. The problem is of a numerical nature. Whereas the signal gets  the
correct inhomogeneous boundary data to propagate it off the evolution
domain, the noise gets the left-over homogeneous data and gets trapped in
exponentially growing modes.

The lesson here is that it is preferable to use Sommerfeld type boundary
conditions, not for physical or mathematical advantage but for numerical
advantage. A Sommerfeld boundary condition doesn't solve all the  problems.
Even though a homogeneous Sommerfeld condition carries energy out of the
evolution domain, it does not in general give the physically correct outer
boundary condition for an isolated system. For a nonlinear system such as
general relativity, one would need an inhomogeneous Sommerfeld condition whose
boundary data could only be determined by matching to an exterior solution. But
numerically a Sommerfeld condition has the great advantage of allowing the
noise to escape through the boundary. Unfortunately, in present numerical
relativity codes, Sommerfeld boundary conditions are inconsistent with the
constraints, which we will discuss later. 

\subsection{Moving boundaries}

Another mechanism by which a reflecting boundary condition can introduce
exponential modes is the repetitive blue shifting off moving boundaries. This
can even happen for a linear wave propagating between two plane boundaries in
Minkowski space. The boundaries can effectively play ping-pong with a wave
packet by arranging the boundary motion to be always toward the packet during
reflection.

Let $\hat x^\alpha =(\hat t,\hat x, \hat y, \hat z)$ be inertia
coordinates, with the reflecting boundaries in the $(\hat x, \hat y)$ plane.
Under reflection, functional dependence of a wave packet traveling in the
positive $\hat z$-direction changes according to
\begin{equation}
     \Phi(\hat t-\hat z)
         \rightarrow \Phi (e^{2\alpha}(\hat t+\hat z)),
\end{equation}       
where the speed of the reflecting wall is
\begin{equation}
      v=  \tanh \alpha.
\end{equation}
After many reflections the energy in the wave grows
by a factor $e^{4\alpha T}$, where $T$ is measured in units
of the crossing-time between reflections.

It is instructive to reinterpret this experiment from a numerical relativity
viewpoint where the spatial coordinates of the boundaries have fixed grid
values. For that purpose, we consider the well-posedness of the IBVP for the
linear wave equation 
\begin{equation}
     g^{\alpha\beta}\partial_\alpha \partial_\beta \Phi =0.
\end{equation}
in a general background spacetime with non-constant metric $g_{\alpha\beta}$,
Again let the evolution domain be the  region $-L\le z\le L$. 

Most of the mathematical literature on well-posedness of the IBVP is based upon
symmetric hyperbolic systems in first derivative
form~\cite{friedrichs,lax,kreiss,secchi2}. We achieve this for the wave
equation by introducing auxiliary variables ${\bf u}=(\Phi,\partial_\alpha
\Phi)$. Standard results then imply a  well posed IBVP for a homogeneous
boundary condition of the matrix form ${\bf Mu}=0$ provided that 
\begin{itemize}

\item the resulting energy flux normal to the boundary has the {\em
dissipative} property
\begin{equation}
          {\cal F}^n({\bf u})\ge 0
\end{equation}
\item ${\bf M}$ has {\em maximal} rank consistent with this
dissipative property and 
\item  ${\bf M}$ is independent of ${\bf u}$.

\end{itemize}

In the present case, the energy flux is determined by the
energy momentum tensor for the scalar field. The flux
normal to the boundary at $z=+L$ is
\begin{equation}
     {\cal F}^n = -n^\alpha (\partial_t \Phi) \partial_\alpha \Phi
\end{equation}
where $n^\alpha =g^{z\alpha}/\sqrt{g^{zz}}$ is the unit normal to the
boundary.

The dissipative condition can be satisfied in a variety of ways.
The choice 
\begin{equation}
       \partial_t \Phi = 0
\end{equation}
leads to a homogeneous Dirichlet boundary condition;
and the choice
\begin{equation}
 n^\alpha \partial_\alpha \Phi = 0 .
\end{equation} 
leads to a homogeneous Neumann boundary condition. Homogeneous
Dirichlet and Neumann boundary conditions are limiting cases for which ${\cal
F}^n=0$, i.e. there is no energy flux across the boundary and signals are
reflected. Between these limiting cases, there is a range of homogeneous
boundary conditions with ${\cal F}^n>0$ of the form  $n^\alpha \partial_\alpha
\Phi + P\partial_t \Phi = 0$, where  $P>0$. Of particular interest is the
Sommerfeld-like case where the derivative lies in an outgoing characteristic
direction. 

The IBVP for the scalar wave equation is well-posed for any of these {\em
maximally dissipative} boundary conditions. Furthermore, by consideration of
the symmetric hyperbolic equation satisfied by ${\bf u}-{\bf q}(x^\alpha)$,
where ${\bf q}$ has explicitly prescribed space-time dependence, the
well-posedness of the IBVP with the homogeneous boundary condition ${\bf Mu}=0$
can be extended to the inhomogeneous form  ${\bf M}({\bf u}-{\bf q}(t,x,y)) =
0$, with freely assigned boundary data  ${\bf q}$. By using such boundary data,
a Neumann or Dirichlet boundary condition can be used to model a wave which is
completely transmitted across the boundary with no reflected component, at
least at the analytic level.

Note the important feature that the free boundary data for the scalar field
consist of one function of three variables in contrast to two functions for
free Cauchy data. As we shall see, this is the major complication in
formulating a constraint preserving boundary condition for a well-posed IBVP in
general relativity.

Now consider the simulation of a linear scalar wave in the flat background
metric which results from the transformation $\hat t  =t$, $\hat x  =x $, $\hat
y  =y$, $\hat z = z+ A \sin\omega t$ from  inertial coordinates $\hat
x^\alpha$.  In these $x^\alpha$ coordinates, the boundaries at $z=\pm L$ are
oscillating relative to the inertial frame, as indicated by the ``shift''
$g^{zt}=-A\omega \cos\omega t$. For our simulation, we prescribe initial data
$\Phi_0=\partial_t \Phi_0=0$ and either the appropriate Neumann or Dirichlet
boundary data $q_{-L}(t)$ and $q_{+L}(t)$ for a wave packet which enters the
boundary at $z=-L$, travels across the domain and leaves  the boundary at
$z=+L$. A second order accurate code would simulate this signal with
$O(\Delta^2)$ error in the grid displacement $\Delta$. Thus $\Phi =O(\Delta^2)$
after the packet has traversed the domain. However, this remnant error gets
homogeneous boundary data. Although, as just discussed,  the normal energy flux
associated with homogeneous Dirichlet or Neumann data vanishes in the rest
frame of the boundary, in the $\hat x^\alpha$ inertial frame the boundary is
moving and the noise can be repeatedly blue shifted, resulting in an
exponential increase of energy. 

One way to eliminate this problem would be to deal with coordinate systems
in which the shift vanishes, at least at the boundary. However, this would
rule out many promising strategies for dealing with binary black holes,
e.g. the  use of co-rotating coordinates or of generalized Kerr-Schild
coordinates.  But especially in a nonlinear problem such as general
relativity, the excitation  of exponential modes can rapidly destroy code
performance.

\section{General Relativity: Harmonic Evolution}
\label{sec:gr}

The previous examples of scalar waves show that even if the underlying analytic
problem is well-posed and even if the numerical simulation converges to the
analytic solution, the existence of exponentially growing modes in the analytic
system can effectively invalidate long term code performance. In general
relativity, coordinate freedom is a further complication which can introduce
rapidly growing modes that are an artifact of gauge pathologies. In order to
illustrate computational problems that are not a trivial consequence of gauge,
we consider the harmonic formulation of Einstein's equations. Although no
coordinates can guarantee complete avoidance of gauge problems, harmonic
coordinates have several advantages for investigating the interface between
numerical and analytic problems in general relativity: 

\begin{itemize}

\item Small number of variables 

\item Small number of constraints (4 harmonic conditions)

\item Einstein's equations reduce to quasilinear wave equations

\item Well-posed Cauchy problem~\cite{bruhat}

\item Symmetric hyperbolic formulation~\cite{fisher}

\item Global asymptotically flat solutions for weak Cauchy data~\cite{rodn} 

\item Well-posed homogeneous IBVP~\cite{szi03}
 
\end{itemize}

A numerical code for evolving Einstein's equations, the Abigel
code~\cite{szi03}, has been based upon a generalized version of harmonic coordinates
satisfying the curved space wave equation 
\begin{equation}
  H^\alpha:= \sqrt{-g}\Box_g x^\alpha= \partial_\mu
        (\sqrt{-g}g^{\mu\nu}\partial_\nu x^\alpha)
   =\tilde H^\alpha (x^\beta,g_{\rho\sigma}),
\end{equation}
where $\tilde H^\alpha$ are harmonic source terms. These harmonic source terms
do not affect any of the analytic results regarding well-posedness but, in
principle, they allow any spacetime to be simulated in some generalized
harmonic coordinate system. The harmonic reduced evolution equations are
written in terms of the metric density $\gamma^{\mu\nu}=\sqrt{-g}g^{\mu\nu}$
whose ten components obey quasilinear wave equations 
\begin{equation}
    \gamma^{\alpha\beta}\partial_\alpha\partial_\beta \gamma^{\mu\nu}
       =S^{\mu\nu},
       \label{eq:harmwaveeq}
\end{equation}
where $S^{\mu\nu}$ contains nonlinear first-derivative terms that do not enter
the principle part. The harmonic conditions $C^\mu: = H^\mu-\tilde H^\mu=0$ are
the constraints on this evolution system which are sufficient to ensure that
Einstein's equations are satisfied. Except where noted, we set $\tilde H^\mu=0$
to simplify the discussion but all results generalize to include nonvanishing
gauge source terms. For details concerning the formulation and its
implementation see Ref.~\cite{szi03}.  

By virtue of the evolution equations,
the harmonic constraints satisfy a homogeneous wave equation of the form
\begin{equation}
    \gamma^{\alpha\beta}\partial_\alpha\partial_\beta C^\mu
    +A^{\mu\alpha}_\beta \partial_\alpha C^{\beta} +B^{\mu}_\beta C^{\beta} =0.
\label{eq:constprop}
\end{equation}
Thus, in the domain of dependence of the Cauchy problem, the solution $C^\mu=0$
is implied by standard uniqueness theorems provided the system is initialized
correctly.

A well-posed evolution system is a necessary but not sufficient ingredient
for building a reliable evolution code. Code performance can be best tested
by  simulating an exact solution and measuring an error norm. The error
should converge to zero in the continuum limit as the grid spacing $\Delta$
shrinks to zero. In testing evolution codes it is desirable to first
eliminate effects of boundary conditions by imposing periodicity in space,
which is equivalent to carrying out the simulation on a 3-torus without
boundary. A suite of toroidal testbeds for numerical relativity has been
developed as part of the Apples with Apples project~\cite{apples,mex1m}. The
convergence and stability of several codes~\cite{apples,palmam,lsu},
including the Abigel code~\cite{apples},
has been demonstrated using this test suite. 

One testbed is the Apples with Apples periodic gauge wave with metric
\begin{equation}
  \label{eq:flatgaugewave4metric}
  ds^2=\Phi(-dt^2 +dz^2)+dx^2+dy^2,
  \label{eq:hgw}
\end{equation}
where
\begin{equation}
       \Phi=1 +A \sin \left( \frac{2 \pi (t-z)}{2L} \right).
\label{eq:sinwave}
\end{equation}
It is obtained from the Minkowski metric
$ds^2 =- d\hat t^2+d\hat x^2+d\hat y^2+d\hat z^2$ by the harmonic coordinate
transformation 
\begin{equation}
  \label{eq:GaugeWave1}
  \begin{array}[c]{r c l}
   \hat t&=& t - \frac {AL}{\pi}\cos \left( \frac{2 \pi (t-z)}{2L} \right), 
      \\
   \hat z&=& z + \frac {AL}{\pi}\cos \left( \frac{2 \pi (t-z)}{2L} \right), 
      \\
     \hat x&=& x,    \\ 
     \hat y&=& y.
  \end{array}
\end{equation}.

\begin{figure}
\centerline{
\begin{psfrags}
\psfrag{xlabel}[ct]{time (crossing times)}
\psfrag{ylabel}[cb]{$||g_{zz}^{ana}-g_{zz}^{num}||_{\infty}$}
\psfrag{Nx=100}[lc]{$N=100$}
\psfrag{Nx=200}[lc]{$N=200$}
\epsfxsize=2.87in\epsfbox{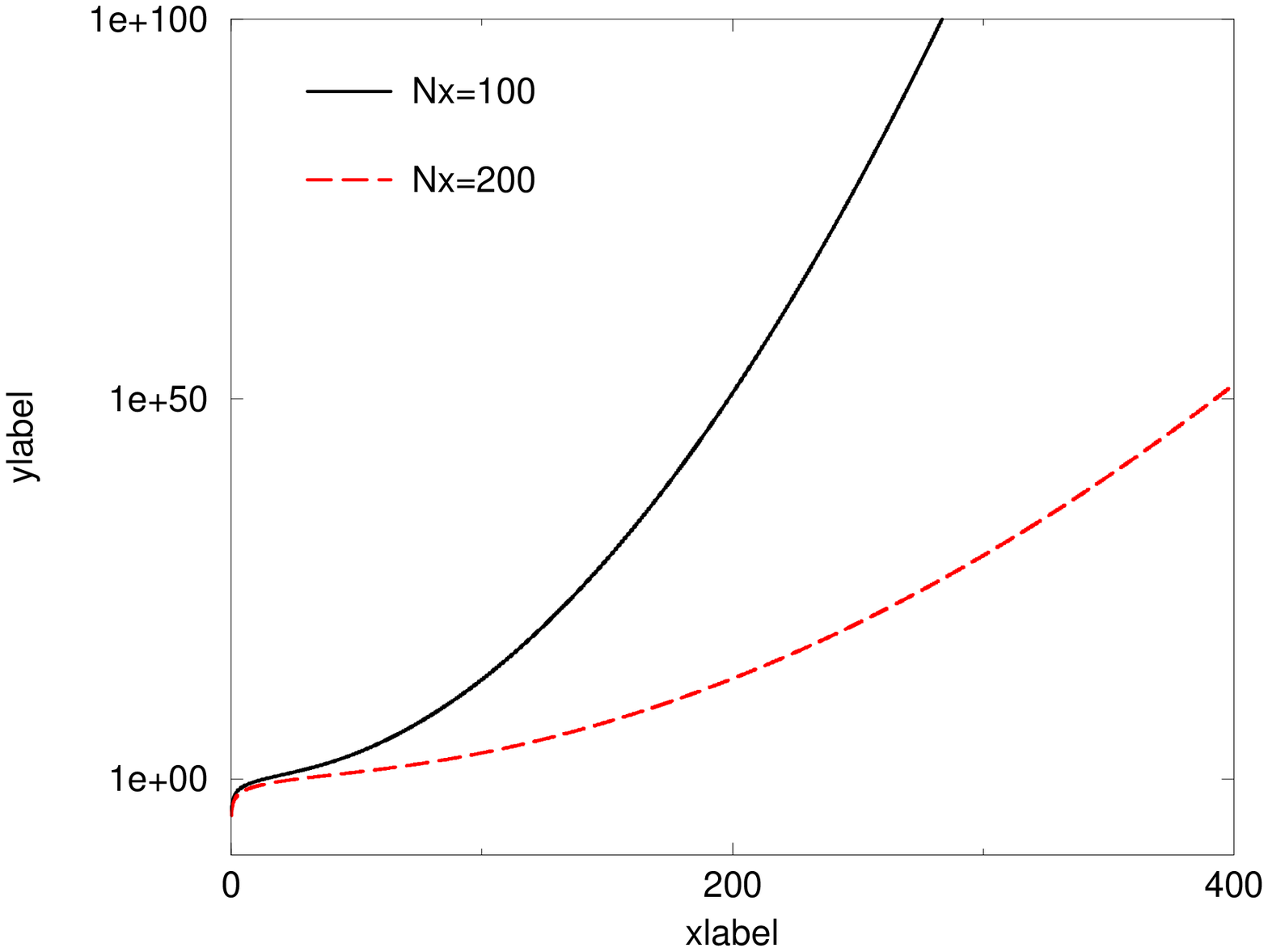}
\end{psfrags}
\begin{psfrags}
\psfrag{xlabel}[ct]{$z$ axis}
\psfrag{gammazz}[lc]{$g_{zz}, t = 0$}
\epsfxsize=2.5in\epsfbox{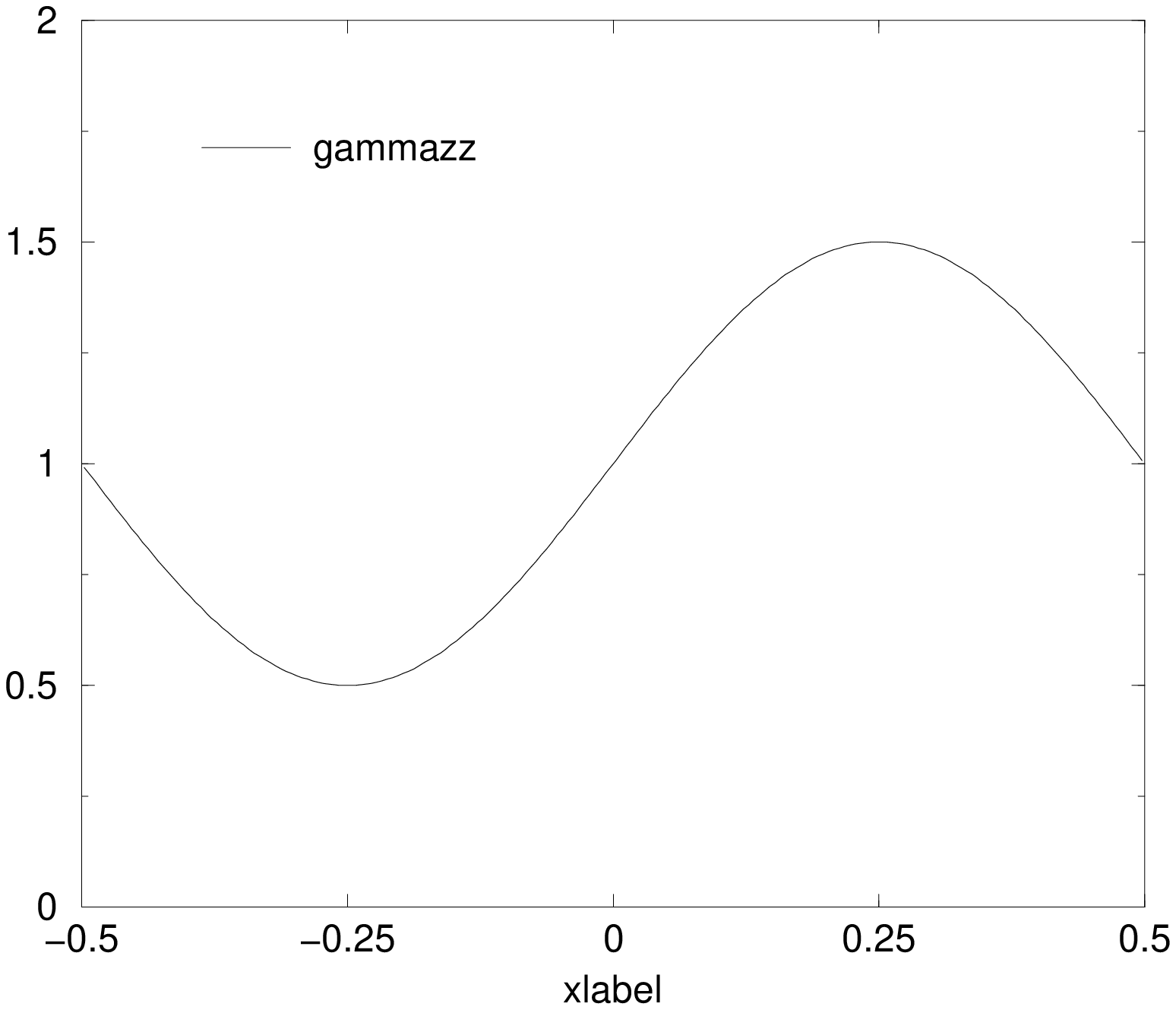}
\end{psfrags}
}
\hspace{0.1in}
\centerline{
\begin{psfrags}
\psfrag{xlabel}[ct]{$z$ axis}
\psfrag{gammazz}[lc]{$g_{zz}, t = 100, N = 100$}
\epsfxsize=2.605in\epsfbox{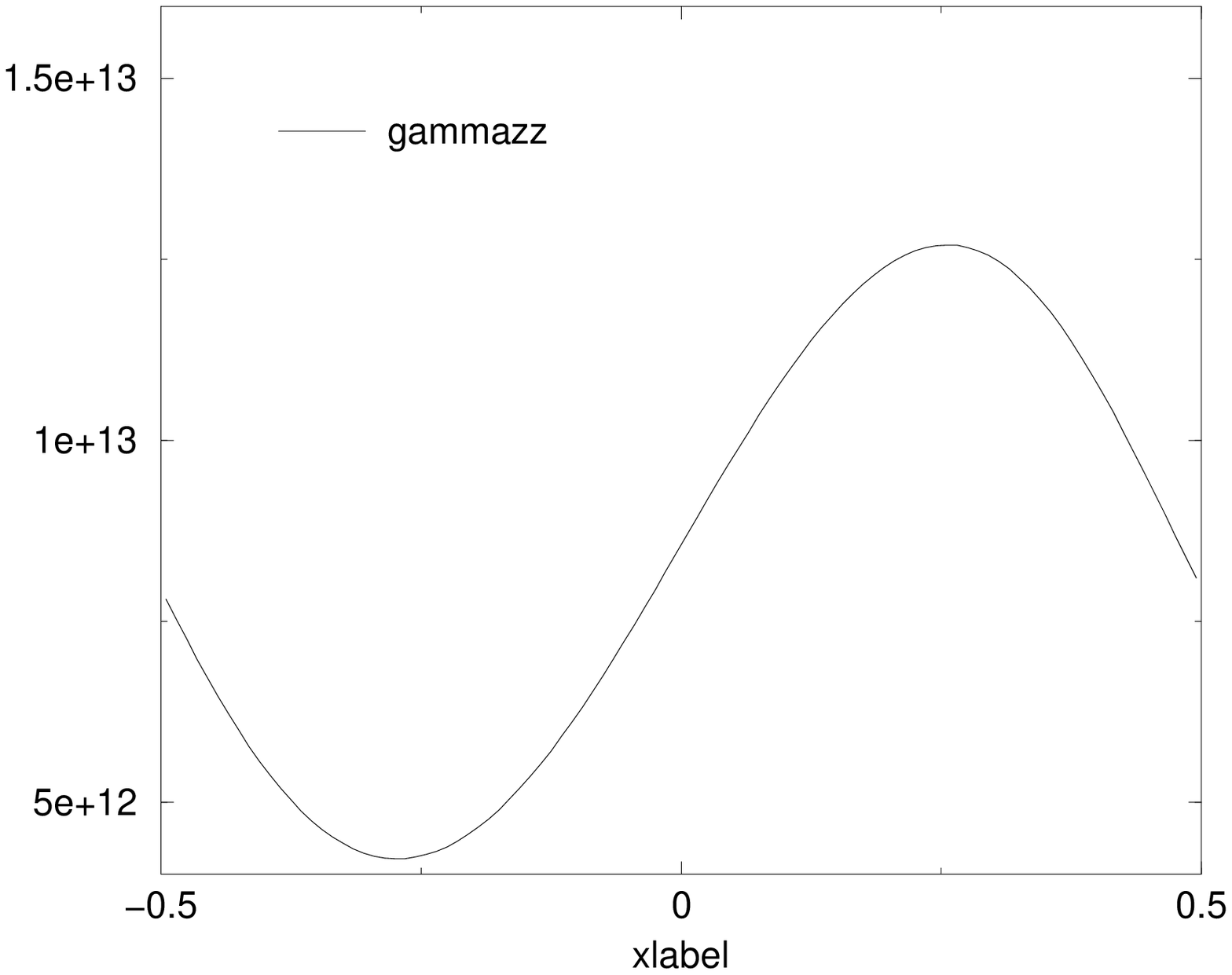}
\end{psfrags}
\begin{psfrags}
\psfrag{xlabel}[ct]{$z$ axis}
\psfrag{gammazz}[lc]{$g_{zz}, t = 100, N = 200$}
\epsfxsize=2.5in\epsfbox{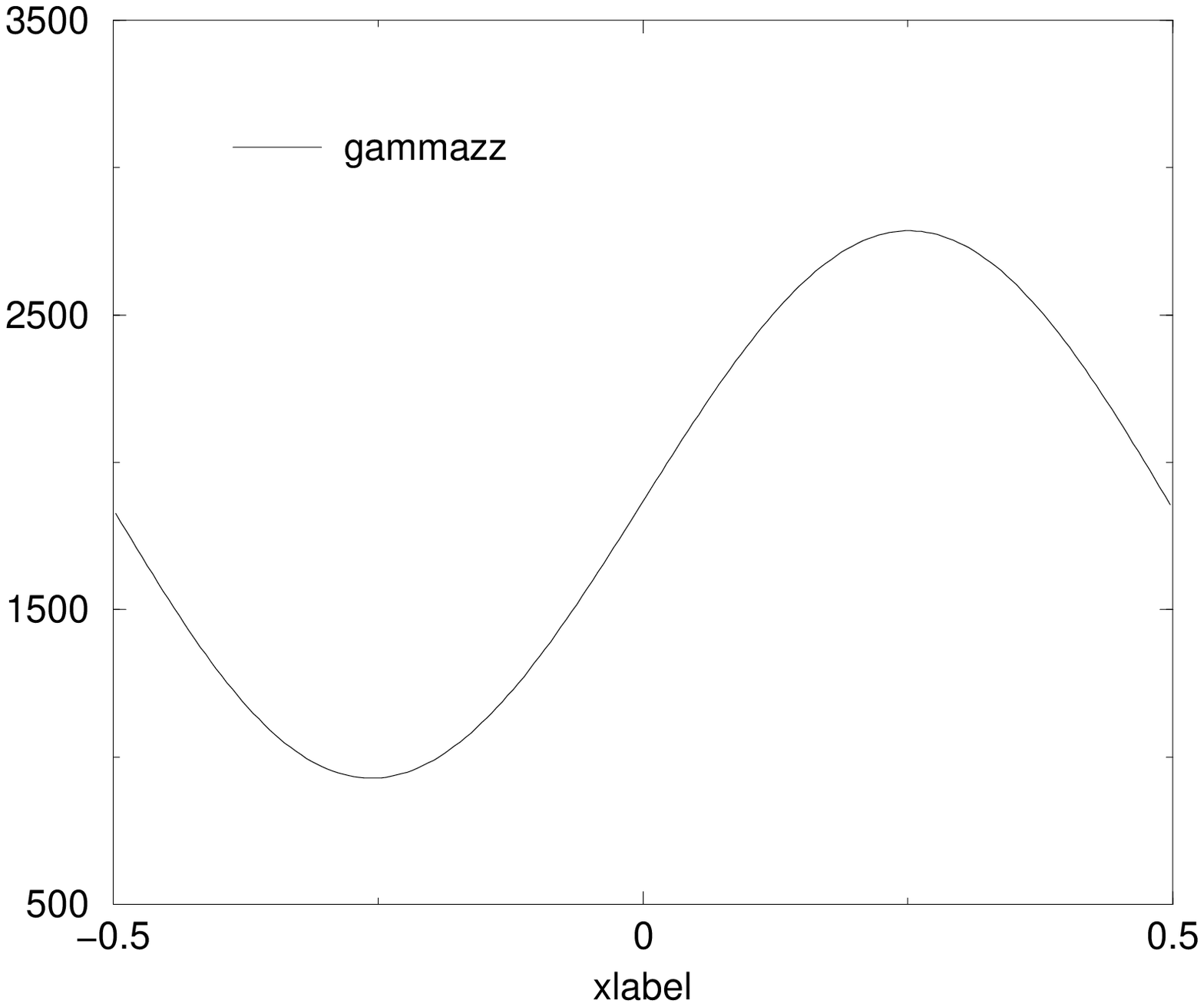}
\end{psfrags}
}
\vspace{0.1in}
\caption{Snap shots of a gauge wave simulation
carried out with an early version of the Abigel
code.
The code is stable and the error converges to
zero at second order in grid spacing $\Delta$
but after a few crossing times the error is
too large to make the results useful.}
\label{fig:gw1D.nolog.tight}
\end{figure}

Figure \ref{fig:gw1D.nolog.tight} shows some snap shots of a gauge wave
simulation carried out with an early version of the Abigel code.
The time dependence of the error shows exponential growth of the form
\begin{equation}
        {\cal E} \sim \Delta^2 \psi(t,z) e^{\lambda t}.
\end{equation}
Here $\psi$ is a well-behaved function which is almost identical in
shape to the signal $\Phi$. As a result, the error cannot be dissipated by
standard techniques for dealing with short wavelength noise.  The exponential
growth originates in exact analogy with our example for the nonlinear scalar
wave equation (\ref{eq:logwave}).  In fact, the metric (\ref{eq:hgw}) is a flat
solution of the Einstein equations in harmonic coordinates for any $\Phi(t,z)$
which satisfies the nonlinear wave equation (\ref{eq:logwave}). The general
solution is $\Phi=e^{f(t-z)+g(t+z)}$. In particular, there are exponentially
growing harmonic gauge waves
\begin{equation}
  ds^2=\Phi_\lambda(-dt^2 +dz^2)+dx^2+dy^2,
\label{eq:lgw}
\end{equation}
where
\begin{equation}
       \Phi_\lambda=e^{\lambda t} 
       \bigg (1 +A \sin \left( \frac{2 \pi (t-z)}{2L} \right)\bigg ),
\end{equation}
which lie arbitrarily close to the gauge wave being simulated.
Thus numerical error inevitably excites exponential modes which
eventually dominate the simulation of the gauge wave. The practicality of
code performance depends on the timescale of this exponential growth.

Although the Abigel code is stable, convergent and based upon a well-posed
initial value problem, like any other code it is subject to the excitation of
exponential modes in the underlying analytic system. Certain numerical
techniques greatly improve its accuracy in simulating the gauge wave:
\begin{itemize}
\item Expressing the equations into {\em flux conservative form},
an idea from computational fluid dynamics which was introduced into
general relativity by the Palma group~\cite{palma}.
\item {\em Summation by parts}, introduced into general relativity by the LSU
group~\cite{lsu}, which at the level of linearized equations leads to energy
estimates for the semi-discrete system of ODE's in time which arise from
spatial discretization.
\item {\em Nonlinear multipole conservation}, which suppresses the
excitation of long wavelength exponential modes by grouping the troublesome
nonlinear terms in a way that enforces global semi-discrete conservation
laws (or approximate conservation laws). 
\end{itemize}
The semi-discrete multipole technique (being introduced here) provides an
excellent example of how analytic insight into the source of a numerical
problem can be used to design a remedy. As we will show, various combinations
of these three techniques lead to dramatic improvement in gauge wave
simulations. Other numerical methods based upon enforcing or damping the
constraints are not crucial for the gauge wave problem but can be important
for simulations of curved spacetimes. 

\begin{figure}
\centerline{
\begin{psfrags}
\psfrag{xlabel}[ct]{time (crossing times)}
\psfrag{ylabel}[cb]{$||g_{zz}^{ana}-g_{zz}^{num}||_{\infty}$}
\psfrag{TIGHT}{TIGHT}
\psfrag{SBP}{SBP}
\psfrag{TIGHT-LOG}{TIGHT-MULT}
\psfrag{SBP-LOG}{SBP-MULT}
\epsfxsize=4in\epsfbox{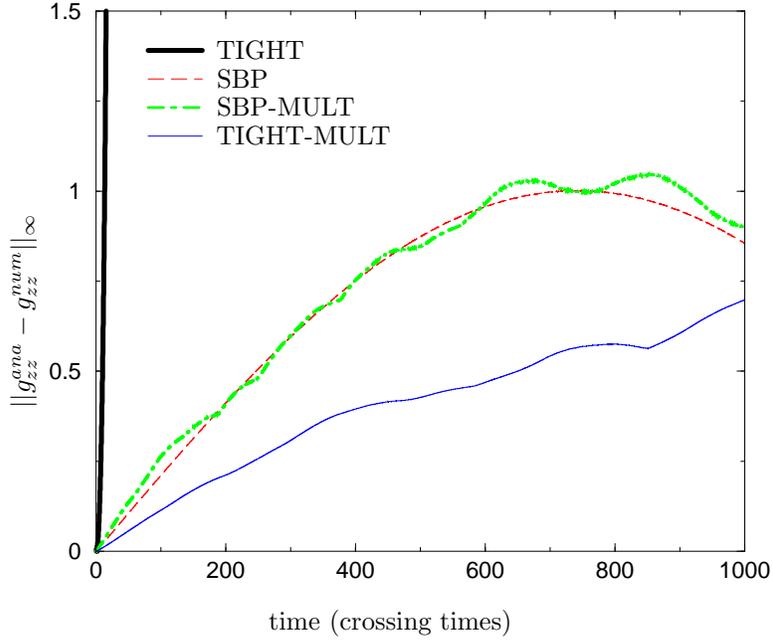}
\end{psfrags}
}
\hspace{0.1in}
\caption{A comparison of the various evolution algorithms used to evolve the
harmonic Einstein equations. In these tests the code evolved flat spacetime in
the gauge defined by Eq.~(\ref{eq:GaugeWave1}), with amplitude $A=.5$.  The
size of the grid was $N=100$, with a time-step of $\Delta t = \Delta z / 4$.
The graph shows the $\ell_\infty$ error norm of the $g_{zz}$ metric component.}
\label{fig:gw1D.cmp}
\end{figure}

An essential ingredient in any code is the method used to approximate
derivatives. The Abigel code treats the quasilinear wave equations
(\ref{eq:harmwaveeq}) as first differential order in time and second order in
space. This allows use of explicit finite difference methods to deal with the
mixed space-time derivatives introduced by the ``shift'' term in the wave
operator while avoiding the artificial constraints that would be introduced by
full reduction to a first order system. On a grid with spacing $\Delta$, the
natural finite difference representation for the the first and second spatial
derivatives are the centered approximations
\begin{equation}
    \partial_z F(z)\rightarrow   D F(z)= 
             \frac {F(z+\Delta) - F(z-\Delta)}{2\Delta}
\end{equation}
and
\begin{equation}
        \partial_z^2 F(z)\rightarrow   D^{(2)} F(z)=
         \frac {F(z+\Delta)-2F(z)+F(z-\Delta)}{\Delta^2}.
\label{eq:tight}
\end{equation} 
These formulae were used in the simulation labeled TIGHT in Fig.
\ref{fig:gw1D.cmp}. Although the code was tested to be stable and convergent
with second order accuracy, the excitation of the exponentially growing mode of
the analytic problem limits accurate simulations to about 10 crossing times on
a reasonably sized grid.

In order to exaggerate nonlinear effects, the simulations shown in Fig.
\ref{fig:gw1D.cmp} were  carried out for a highly nonlinear gauge wave with
amplitude $A=.5$, on a scale where the metric is singular for $A=1$. (The
standard Apples with Apples tests specify amplitudes of $A=.01$ and $A=.1$.)
Problems with exponential modes do not appear for small amplitudes
simulations in the linear domain. One contributing factor to the exponential
growth is that the tight 3-point stencil (\ref{eq:tight}) for the second
derivative does not lead to an exact finite difference representation of the
integration by parts rule necessary to establish energy conservation, which
is the main idea behind the summation by parts (SBP) method. But this is only
part of the story since standard SBP techniques only apply to linear systems.

It is instructive to examine how these ideas extend to the second derivative
form of the nonlinear wave equation (\ref{eq:logwave}) which underlies the
gauge wave problem. This will illustrate in a simple way how flux conservative 
equations, SBP and multipole conservation can combine to suppress excitation
modes in the analytic problem. The model scalar problem is effective in
isolating the difficulties underlying a full general relativistic code, in
addition to allowing efficient computational experimentation.

We carry out the analysis for waves traveling with periodic boundary
conditions in one spatial dimension. The extension to three dimensions is
straightforward but notationally more complicated. The theory regarding
well-posedness of hyperbolic systems is based upon the principle part of the
equations. For that reason, we   first consider the linear wave equation
\begin{equation}
    \partial_\alpha\partial^\alpha\Phi 
        =-\partial_t^2 \Phi + \partial_z^2 \Phi=0.
	\label{eq:lw} 
\end{equation}    
The energy associated with this wave can be related to the conserved
integral
\begin{equation}
      {\cal I}= [\Phi_1,\Phi_2]=\oint (\Phi_1 \partial^\mu \Phi_2
                 -\Phi_2 \partial^\mu \Phi_1) dV_\mu
\end{equation}
by choosing
$\Phi_1=\Phi$ and $\Phi_2=\partial_t \Phi$.
For the case of periodic boundary conditions on the interval $0\le z\le L$,
\begin{equation}
    {\cal I} =\int_0^L \bigg ( (\partial_t \Phi)^2
                        - \Phi \partial_t^2 \Phi \bigg )dz.
\end{equation}
The integration by parts by parts rule
\begin{equation}
      \int_0^L \bigg( -\partial_z(\Phi \partial_z \Phi)
      + (\partial_z\Phi)^2 + \Phi \partial_z^2 \Phi  \bigg )dz=0,
\label{eq:intleib}
\end{equation}
applied to a periodic interval,
then supplies the key step in using the wave equation to relate ${\cal I}$
to the positive definite energy
\begin{equation}
   {\cal I}= {\cal E} =\int_0^L \bigg ( (\partial_t \Phi)^2
                +(\partial_z \Phi)^2  \bigg ) dz.
\end{equation}

In order to obtain a discrete version of the integration by parts identity
(\ref{eq:intleib}), we introduce a uniform grid $z_i$, $0\le i \le N$, with
spacing $\Delta$ and represent 
\begin{equation}
       \int_0^L F dz \rightarrow \Delta\sum_0^N f_{i+1/2}
\end{equation}
where
\begin{equation}
          f_{i+1/2}=\frac{F(z_i)+F(z_{i+1})}{2}.
\end{equation}
In addition we represent derivatives at the midpoints by the centered
approximation 
\begin{equation}
         \partial_z F(z_i+\Delta/2) 
         \rightarrow f^\prime_{i+1/2}=  \frac{F(z_{i+1})-F(z_i)}{\Delta}
\end{equation}
so that periodic boundary conditions imply
\begin{equation}
       \int _0^L \partial_z F dz  
         \rightarrow  \Delta\sum_0^N f^\prime_{i+1/2} =F|_0^L =0.
\end{equation}
This ensures the semi-discrete monopole conservation law
\begin{equation}
       \partial_t^2 \oint \Phi dz \rightarrow 0,
       \label{eq:moncons}
\end{equation}
which results from the flux conservative form of Eq. (\ref{eq:lw}). Equation
(\ref{eq:moncons}) controls growth of the spatial average of $\Phi$ but not of
its non-constant spatial Fourier components which measure its gradient. 

Energy estimates control the growth of the gradient of $\Phi$.
With the above definitions, it is straightforward to check that
\begin{equation}
     \partial_z(FG) - F\partial_z G -G\partial_z F \rightarrow 0.
\label{eq:leib}
\end{equation}
As a result, the semi-discrete version of the integral identity
(\ref{eq:intleib}) is satisfied if the second derivative term is represented as
a product of first derivatives. For the linear wave equation this results in
the semi-discrete conservation law
\begin{equation}
       \partial_t {\cal E} \rightarrow 0.
\end{equation}

\begin{figure}
\centerline{
\begin{psfrags}
\psfrag{xlabel}[ct]{time (crossing times)}
\psfrag{ylabel}[cb]{$||\Phi^{ana}-\Phi^{num}||_{\infty}$}
\psfrag{TIGHT}{TIGHT}
\psfrag{SBP}{SBP}
\psfrag{TIGHT-LOG}{TIGHT-MON}
\psfrag{SBP-LOG}{SBP-MON}
\epsfxsize=4in\epsfbox{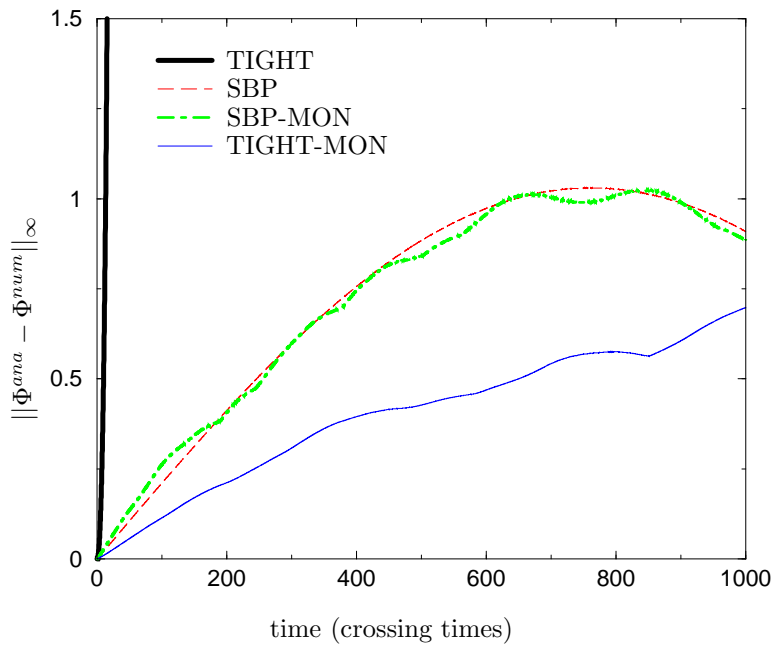}
\end{psfrags}
}
\vspace{0.1in}
\caption{A comparison of the various evolution algorithms used to evolve the
nonlinear wave equation (\ref{eq:logwave}). The tests are based on the sine
wave solution (\ref{eq:sinwave}), with amplitude $A=.5$, simulated on a grid of
$N=100$ points with a time-step of $\Delta t = \Delta z / 4$. The graph shows
the $\ell_\infty$ norm of the error.}
\label{fig:scalar}
\end{figure}

In order to implement SBP in a code such as the Abigel code,
which is second differential order in space with
the fields represented by their values on grid points, and not on 
mid-points, the above results can be applied by treating the mid-points for
even numbered grid points as the odd-numbered grid points, and vice versa. 
This results in the widened finite difference representation for the second
derivative
\begin{equation}
   \partial_z^2 F(z)\rightarrow   D^2 F (z)=DD F(z)=
              \frac {F(z+2\Delta)-2 F(z)+F(z-2\Delta)}{4\Delta^2},
\label{eq:wide}
\end{equation}
as opposed to the tight stencil (\ref{eq:tight}). Figure \ref{fig:scalar}
shows the remarkable improvement in long term accuracy obtained in the
simulation of a non-linear wave satisfying Eq. (\ref{eq:logwave}).
(Numerical dissipation has been used to damp short wavelength instabilities
triggered by the loose coupling between even and odd grid points.) The
curves labeled TIGHT are obtained using the standard stencil
(\ref{eq:tight}). They show exponential growth on a scale of $\approx 10$
grid crossing times. The curves labeled SBP are obtained using the stencil
(\ref{eq:wide}) consistent with SBP. This change of stencil suppresses
growth of long wavelength exponential modes so that accurate simulations of
$\approx 1000$ crossing times are possible. 

Summation by parts only has general applicability to linear equations, although
the technique extends in an approximate sense to the nonlinear domain. Other
approaches can also be successful for nonlinear problems, especially if  the
troublesome nonlinear terms can be identified. In the case of the nonlinear
equation (\ref{eq:logwave}), these terms can be incorporated in the principle
part by reformulating the equations in the flux conservative form
\begin{equation}
    \partial_\alpha \bigg(  \eta^{\alpha\beta}\frac{1}{\Phi}
          \partial_\beta \Phi \bigg) =0,
\label{eq:lwave}
\end{equation}
with the subsequent reduction reduction to the first order in time system
\begin{eqnarray}
     \partial_t \Phi &=& \Phi Q \\
     \partial_t Q &=& \partial_z (\frac{1}{\Phi} \partial_z \Phi).
\end{eqnarray}
Many choices of spatial discretization of this flux conservative system  
lead to an {\em exact} semi-discrete version of the monopole conservation
law
\begin{equation} 
            \partial_t \int_0^L Q =0.
\end{equation}
We consider the two choices
\begin{equation}
     \partial_z (\frac{1}{\Phi} \partial_z \Phi)|_i \rightarrow 
      \frac{1}{2\Delta} \bigg( \frac {1}{\Phi} \Phi^\prime \bigg)_{i+1} 
       -  \frac{1}{2\Delta} \bigg(\frac {1}{\Phi} \Phi^\prime \bigg)_{i-1}
\label{eq:widenl} 
\end{equation}    
and
\begin{equation}
     \partial_z (\frac{1}{\Phi} \partial_z \Phi)|_i \rightarrow 
      \frac{1}{\Delta} \bigg( \frac {1}{\Phi} \Phi^\prime \bigg)_{i+1/2} 
       -  \frac{1}{\Delta} \bigg(\frac {1}{\Phi} \Phi^\prime \bigg)_{i-1/2}. 
\label{eq:tightnl}
\end{equation}  
As a result of either of these discretizations, the initial data determine the
conserved value of the monopole moment $\int_0^L Q dz$ and the excitation of
the exponential modulation (\ref{eq:expt}) is thereby frozen out of the
numerical evolution. In this way a tight 3-point stencil (\ref{eq:tightnl}) can
be used, as opposed to the wide 5-point stencil (\ref{eq:widenl}) (and the
concomitant numerical dissipation) required by SBP for the second order system.
The curve labeled TIGHT-MON in Fig. \ref{fig:scalar} shows how long term
accuracy is dramatically enhanced by this technique, without use of numerical
dissipation. The curve labeled SBP-MON shows that, in this case, no additional
improvement is gained when monopole conservation is combined with SBP.

These numerical techniques introduced for the model scalar problem were
formulated in a way that could be readily taken over to the gravitational case.
Although the Einstein equations can neither be linearized by taking the
``logarithm of the metric'' nor written in a completely flux conservative form
analogous to (\ref{eq:lwave}), there are various ways to group derivatives
which decouple the Jacobean transformation that generates the exponential mode
in the gauge wave metric (\ref{eq:lgw}). One example is the grouping  
\begin{equation}
     g^{\alpha\mu}\partial^\rho g_{\mu\beta} =
          (\delta^\alpha_t\delta^t_\beta +\delta^\alpha_z\delta^z_\beta)
                 \frac{1}{\Phi_\lambda}\partial^\rho\Phi_\lambda ,
\end{equation}
for which expression of the principle part of the Einstein equation
in the form $\partial_\rho (g^{\alpha\mu}\partial^\rho g_{\mu\beta})$
leads to the semi-discrete conservation laws
\begin{equation}
        \partial_t \int_0^L {g^{\alpha\mu}\partial^t g_{\mu\beta}}dz
            \rightarrow 0
\end{equation}
for the gauge wave. The conserved quantities are comprised of multipoles of
monopole (the spatial trace) and quadrupole (the trace-free part) type.

The advantage of enforcing these conservation laws is exhibited in Fig.
\ref{fig:gw1D.cmp}. Comparison of Figs. \ref{fig:scalar} and
\ref{fig:gw1D.cmp}  shows that SBP and multipole conservation lead to almost
identically beneficial results in simulating the gauge wave as in simulating
the nonlinear scalar wave. The standard 3-point stencil (TIGHT) again excites
exponentially growing error on the order of 10 crossing times but accurate
evolutions for over 1000 crossing times are attained either with SPB or with a
3-point stencil embodying multipole conservation (TIGHT-MULT).

\begin{figure}
\centerline{
\begin{psfrags}
\psfrag{xlabel}[ct]{time (crossing times)}
\psfrag{ylabel}[cb]{$||g_{zt}^{ana}-g_{zt}^{num}||_{\infty}$}
\psfrag{TIGHT}{TIGHT}
\psfrag{SBP}{SBP}
\psfrag{TIGHT-LOG}{TIGHT-MULT}
\psfrag{SBP-LOG}{SBP-MULT}
\epsfxsize=4in\epsfbox{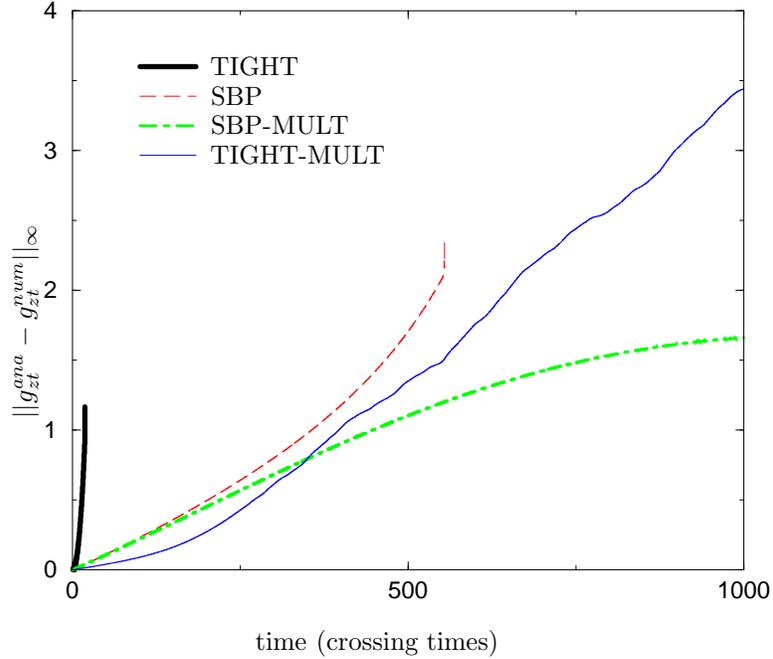}
\end{psfrags}
}
\hspace{0.1in}
\caption{A comparison of the various evolution algorithms used to evolve the
harmonic Einstein equations.  In these tests the code evolved the gauge wave
metric with shift defined in Eq.~(\ref{eq:shgw}), with amplitude $A=.5$. The
size of the grid was $N=100$, with a time-step of $\Delta t = \Delta z / 4$.}
\label{fig:gw1Dsh.cmp}
\end{figure}

Whether objectionable gauge modes can be decoupled so effectively in a more
general problem is an interesting question. However, all the above numerical
techniques, which lead to excellent code performance for the gauge wave, are
of a universal nature that can be adopted for the simulation of a general
spacetime by a general code. Since most simulations contain a weak field
region, such as the far field region outside a black hole, these techniques
might in fact be necessary in order to avoid excitation of local versions of
exponential Minkowski gauge modes. Figure \ref{fig:gw1Dsh.cmp} shows how
these methods extend to the challenging simulation of the gauge
wave with shift 
\begin{equation}
   ds^2= -(1-A \sin\alpha)dt^2+2A \sin\alpha dtdz
         +(1+A \sin\alpha)dz^2 + dx^2 +dy^2,
	 \label{eq:shgw}
\end{equation}
with $\alpha =\pi(t-z)/L$. The simulation was carried out with periodic
boundary conditions and amplitude $A=.5$, so that the grid has an effective
velocity of half the speed of light. Again there are exponentially growing
gauge waves,
\begin{equation}
   ds_\lambda^2= -(e^{\lambda t}-A \sin\alpha)dt^2+2A \sin\alpha dtdz
         +(e^{\lambda t}+A \sin\alpha)dz^2 + dx^2 +dy^2
	 \label{eq:shlgw}
\end{equation}
(for arbitrary $\lambda$), which satisfy these boundary conditions. These will
trigger a numerical instability unless their excitation is controlled by a
conservation law on the semi-discrete system. Remarkable improvement in long
term performance is achieved by implementing either SBP or the multipole
algorithm.

These examples show what must be done, beyond having a stable, convergent code,
in order to achieve accurate long term simulations. Exponential modes
undoubtedly arise in a wide variety of systems with the examples presented here
just the tip of the iceberg. Short wavelength modes arising from discretization
error can be suppressed by numerical dissipation. The long wavelength modes
exist in the analytic problem. This raises some key questions: Are there
geometric clues to identify the origin of such long wavelength exponential
modes? What numerical or analytic techniques can be used to suppress them?

\section {The Harmonic IBVP}

Given an evolution code on the 3-torus which is based upon a well-posed Cauchy
problem for Einstein's equations and which is free of all numerical problems,
several things can go wrong in extending the evolution to include a boundary.
On the analytic side, the imposition of the boundary condition can be ill-posed
or it can lead to violation of the constraints or it can introduce
exponentially growing modes. On the numerical side, the finite difference
implementation of the boundary condition can be unstable or inaccurate. On the
physical side, the correct boundary data representing radiation (or the absence
of radiation) entering the system might not be known or it might not be
possible to extract the waveform of the outgoing radiation. 

Here we examine the analytical and numerical issues for the
harmonic IBVP. The reduced evolution system consists of the 
quasilinear wave equations (\ref{eq:harmwaveeq}). Our discussion for
nonlinear scalar waves show that the IBVP for this system is well-posed
for any maximally dissipative boundary conditions, e.g.
Dirichlet, Sommerfeld or Neumann.

Next consider the harmonic constraints $C^\mu$. They satisfy the homogeneous
wave equation (\ref{eq:constprop}). Thus we can formulate a well-posed IBVP for
the propagation of the constraints by imposing a maximally dissipative boundary
condition. Then, given that the constraints and their time derivative are
satisfied by the initial data and that the constraints have homogeneous
boundary data, the uniqueness of the solution to the constraint propagation
equations would imply that the constraints be satisfied in the domain of
dependence of the IBVP. However, consistency between the boundary conditions
for the evolution variables and the homogeneous boundary conditions for the
constraints is not straightforward to arrange.

For example, consider evolution in the domain $z<0$ with boundary at $z=0$. In
the tangential-normal 3+1 decomposition $x^\mu=(x^a,z)$ intrinsic to the 
boundary, a homogeneous Dirichlet condition on the
constraints takes the explicit form
\begin{eqnarray}
   C^z&=&\partial_a \gamma^{za} +\partial_z \gamma^{zz}=0 \\
     C^a&=&\partial_b \gamma^{ab} +\partial_z \gamma^{az}=0.
\end{eqnarray}
A naive attempt to satisfy these conditions by boundary data on the evolution
variables would involve assigning  both Dirichlet (tangential) and Neumann
(normal) conditions to $\gamma^{az}$, which would be an inconsistent boundary
value problem.

One way to impose consistent constraint preserving boundary conditions is based
upon the well-posedness of the Cauchy problem. Consider smooth Cauchy data
which is locally reflection symmetric with respect to the boundary at $z=0$.
Then in some neighborhood$-L<z<L$ of the hypersurface $z=0$ the Cauchy problem
is well-posed. On $z=0$, the local reflection symmetry implies that the
evolution equations satisfy
\begin{eqnarray}
     \gamma^{za}&=&0 \nonumber \nonumber \\
      \partial_z \gamma^{zz}&=& 0 \nonumber \\
       \partial_z \gamma^{ab} &=& 0 
\label{eq:evbc}        
\end{eqnarray}
and that the constraints satisfy
\begin{eqnarray}
     C^{z}&=& 0 \nonumber\\
    \partial_z C^a &=& 0.
\label{eq:cbc}
\end{eqnarray}
It is straightforward (although algebraically complicated) to show  for the
harmonic IBVP that the combination of Dirichlet and Neumann boundary conditions
(\ref{eq:evbc}) implies that the constraints satisfy the homogeneous boundary
conditions (\ref{eq:cbc}). Thus (\ref{eq:evbc}) provide homogeneous constraint
preserving boundary conditions for a well-posed harmonic IBVP.

Well-posedness of the IBVP extends to the case of ``small'' boundary data, of
the form ${\bf M(u -q}(x^a){\bf )}=0$ discussed in Sec. \ref{sec:gr}, in the
sense that the prescribed data ${\bf q}$ is linearized off a solution with
homogeneous boundary data. However, the available mathematical theorems  do not
guarantee well-posedness for finite boundary data. We describe below the major
issues regarding constraint preserving inhomogeneous boundary conditions for
the harmonic IBVP. For further details, see Ref.~\cite{szi03}. 

Part of the inhomogeneous boundary data which generalize (\ref{eq:evbc})
are associated with the gauge freedom corresponding to a boundary version
of the ``shift''. By a harmonic coordinate transformation it is alway possible
to set
\begin{equation} 
      \gamma^{za}= q^a(x^b)\gamma^{zz}
\end{equation}
at the boundary, where $q^a$ is freely prescribed data. The unit normal $N^\mu$
to the boundary then defines the normal derivative
\begin{equation}
   \partial^n:=\frac{1}{N^z}N^\mu \partial_\mu= \partial_z +q^a\partial_a
\end{equation}
entering the Neumann boundary data, $q^{zz}=\partial^n \gamma^{zz}$ and 
$q^{ab}=\partial^n \gamma^{ab}$, which complete the inhomogeneous version of
(\ref{eq:evbc}). 

The boundary data ${\bf q}= (q^a,q^{zz},q^{ab})$ can be freely prescribed
in a well-posed IBVP for the reduced evolution system but they must be
restricted to satisfy (\ref{eq:cbc}) in order to ensure that the
constraints are satisfied. The condition  $C^z= 0$ requires 
\begin{equation}
            q^{zz}= -\partial_a q^a \gamma^{zz} .
\end{equation}
When the boundary shift $-q^a$ is nonzero, the second condition in
(\ref{eq:cbc}) must be restated in the form $\partial^n C_a= 0$, because the
derivative $\partial_z$ is no longer in the normal  direction to the boundary.
This condition is a restriction on the data $q^{ab}$, which are closely related
to the extrinsic curvature $K^{ab}$ of the boundary. It requires that
\begin{equation}
    \sqrt{-h}D_b (K_a^b-\delta_a^b K)+\sqrt{g^{zz}}K_{ab}C^b
   - \frac{g^{zz}}{2} C_b\partial_a q^b=0 ,
\end{equation}
where $h_{ab}$ and  $D_a$ are the metric and connection intrinsic to the
boundary. This equation can be recast as a symmetric hyperbolic boundary
system which determines the 6 pieces of Neumann data $q^{ab}$ in terms of
3 free functions, the free (gauge) data $q^a$ {\em and} the boundary
values of $\gamma^{zz}$, $\gamma^{ab}$ and $\partial_z \gamma^{za}$.  {\em
Solutions of reduced equations with this boundary data necessarily satisfy
the constraints}. Unfortunately, the appearance of the quantities 
$\gamma^{zz}$, $\gamma^{ab}$ and $\partial_z \gamma^{za}$ complicates the
well-posedness of the constrained IBVP since these quantities cannot be
freely specified but must be determined in the course of the evolution.

Formally, the constraint preserving boundary data have the functional
dependence ${\bf q}= {\bf q}({\bf u},x^a)$, which involves  evolution
variables ${\bf u}$ whose boundary values cannot be freely prescribed. This
complication has its geometric origins in the fact that the boundary data
(gauge quantities and extrinsic curvature) do not include the intrinsic
metric, as in the case of Cauchy data. Because of the dependence of the
constraint preserving boundary data on ${\bf u}$, the available theorems
regarding well-posedness only apply to perturbations of homogeneous data,
where the background values of ${\bf u}$ can be explicitly determined.

\begin{figure}
\begin{psfrags}
\psfrag{Gzz120}[lb]{\small{$|\gamma^{zz}_e|_\infty, A=10^{-1}, 120^3$}}
\psfrag{Gzz80}[lb]{\small{$|\gamma^{zz}_e|_\infty, A=10^{-1}, 80^3$}}
\psfrag{Gzz80lin}[lb]{\small{$|\gamma^{zz}_e|_\infty, A=10^{-3}, 80^3$}}
\psfrag{Hn120}[lb]{\small{$|H|_\infty, A=10^{-1}, 120^3$}}
\psfrag{Hn80}[lb]{\small{$|H|_\infty, A=10^{-1}, 80^3$}}
\psfrag{time}[cb]{\small{$t$ (crossing times)}}
\centerline{\epsfxsize=5in\epsfbox{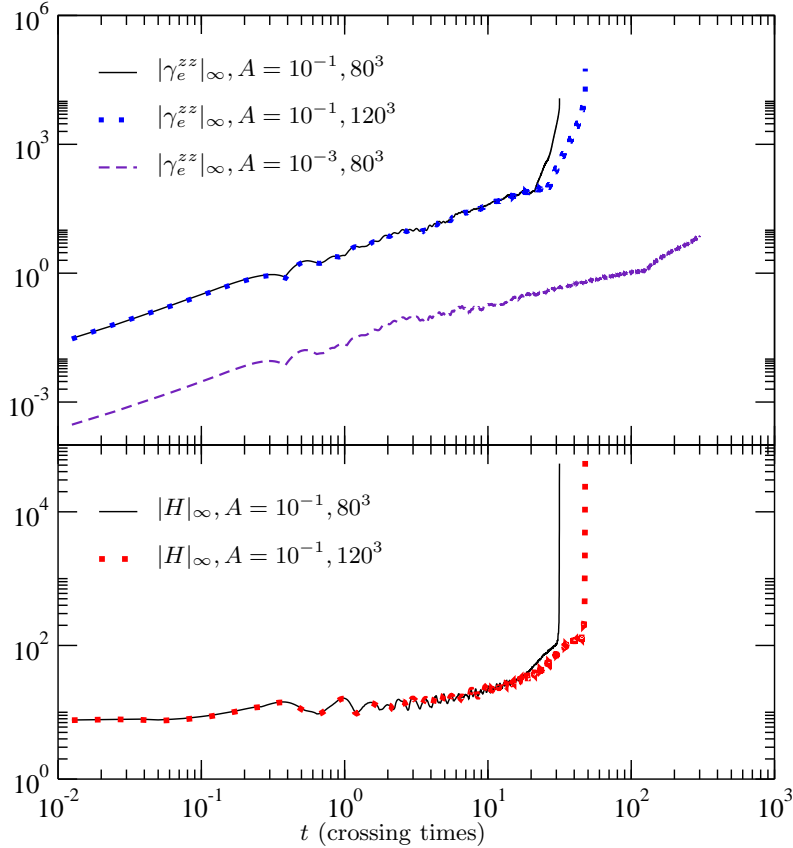}}
\end{psfrags}
\caption{The $\ell_\infty$ norm of the finite-difference error
$\gamma^{zz}_e = \gamma^{zz}_{ana} - \gamma^{zz}_{num}$,  rescaled  by a
factor of $1/\Delta^2$,  for a gauge-wave. The tests were carried
out with an early version of the Abigel code before
semi-discrete conservation laws were incorporated. The upper two (mostly
overlapping) curves demonstrate convergence to the analytic solution for a
wave with amplitude $A = 10^{-1}$ % evolved for 30 crossing times, with
gridsizes $80^3$ and $120^3$. We also plot $|H|_\infty$, the $\ell_\infty$
norm of $\sqrt {(H^t)^2+\delta_{ij}H^i  H^j}$, to demonstrate that
convergence of the harmonic constraints is enforced by the boundary
conditions. The lower curve represents evolution of the same gauge-wave
with $A = 10^{-3}$ for $300$ crossing times, with gridsize $80^3$.}
\label{fig:oldcode}
\end{figure}

These constraint preserving boundary conditions have been implemented in
the Abigel code. Test simulations of the IBVP for the shifted gauge wave
(\ref{eq:shgw}) were carried out by opening one face of the 3-torus to
form a $T^2\times [0,1]$ manifold with boundary. Figure~\ref{fig:oldcode}
shows the results reported for an early version of the code \cite{szi03}.
The graphs indicate stability and and convergence but there is also a
growing error which eventually leads to a nonlinear instability. One
underlying cause of this error growth is the continuous blue shifting off
the moving boundaries, as discussed in Sec.~\ref{sec:waves}. However,
these tests were carried out before semi-discrete conservation laws were
incorporated into the evolution algorithm so that a better understanding
of the error must await future test runs.  

\section{Sommerfeld Alternatives}
\label{sec:somm}

The examples presented here indicate a computational advantage in formulating
boundary conditions in a manner such that numerical noise can propagate off the
grid for the case of homogeneous boundary data. To date, there exists only one
well-posed formulation of the IBVP for general relativity that allows this type
of generalized Sommerfeld boundary condition. This is the Friedrich-Nagy
formulation~\cite{Friedrich98} based upon a formulation of Einstein's equations
in which an orthonormal tetrad, the connection and the Weyl curvature are
treated as evolution variables. The gauge freedom in the theory is adapted in a
special way to the boundary so that boundary conditions need only be imposed on
the curvature variables. The critical feature of the formalism is that the
constraints propagate tangential to the boundary. This allows the
well-posedness of the IBVP for the reduced evolution system to be extended to
the fully constrained system.  Unfortunately, this formulation has not yet been
implemented as a numerical code, partially because of its analytic complexity
and partially because it would require some infrastructure beyond that existing
in most present codes. 

An important issue is whether this success of the Friedrich-Nagy system in
handling a Sommerfeld boundary condition is limited to formulations that
include the tetrad or the curvature among the basic evolution variables. In
linearized gravitational theory, there is a simple variant of the harmonic
formulation that has a well-posed IBVP, admits a Sommerfeld boundary condition
and has been successfully implemented computationally~\cite{hbdry}. The
nonlinear counterpart consists of the evolution system 
\begin{eqnarray}
    \gamma^{\alpha\beta}\partial_\alpha\partial_\beta \gamma^{ij}
       &=& S^{ij}, \\
        H^\alpha: =  \partial_t \gamma^{t\alpha}
	     + \partial_j \gamma^{j\alpha}
	    &=& \hat H^\alpha(x,\gamma),
\end{eqnarray}
comprised of the wave equations (\ref{eq:harmwaveeq}) for the six spatial
components $\gamma{ij}$ and propagation equations for the time components
$\gamma^{t\alpha}$. Alternatively, the propagation equations could be
reformulated  as
\begin{equation}
        \partial_t H^\alpha=\partial_t \hat H^\alpha
\end{equation}
in order to make the evolution system uniformly second differential order.
Well-posedness of the nonlinear Cauchy problem does not follow in any
direct way from standard theorems. An analysis of the principle part shows
that this naive harmonic system is only weakly hyperbolic, which opens the
door for lower derivative terms to produce instabilities~\cite{kreissort}.

\begin{figure}
\centerline{
\begin{psfrags}
\psfrag{xlabel}[ct]{time (crossing times)}
\psfrag{ylabel}[cb]{$\||ham||_{\infty}$}
\epsfxsize=4in\epsfbox{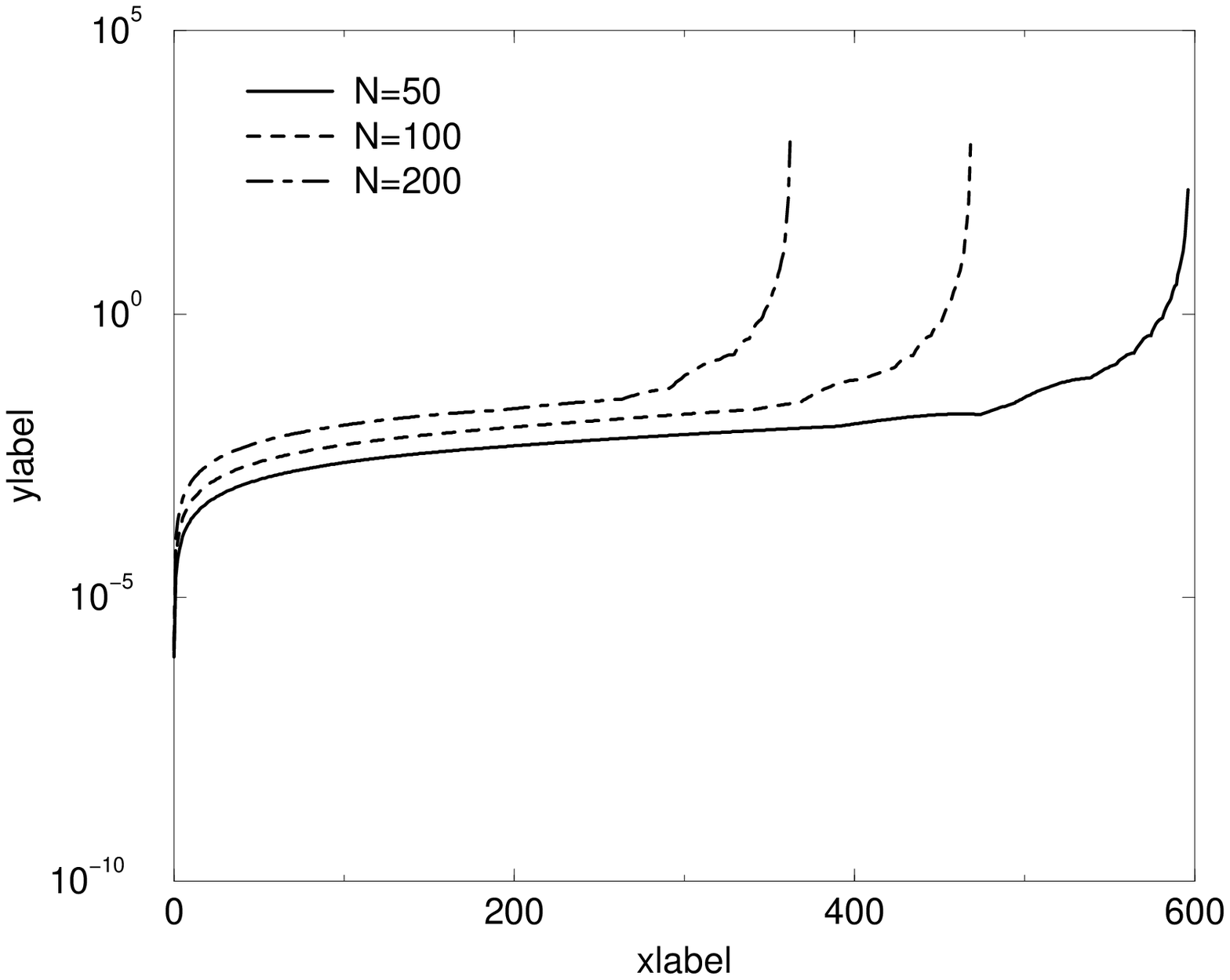}
\end{psfrags}
}
\hspace{0.1in}
\caption{The robust stability test for the weakly hyperbolic harmonic system.
The $\ell_\infty$ norm of the Hamiltonian constraint is plotted on a
linear-logarithmic scale. All specifications are in accord with the Apples With
Apples test.} 
\label{fig:robustH.cmp}
\end{figure}

\begin{figure}
\centerline{
\begin{psfrags}
\psfrag{xlabel}[ct]{time (crossing times)}
\psfrag{ylabel}[cb]{$||g_{zz}^{error}||_{\infty}$}
\epsfxsize=4in\epsfbox{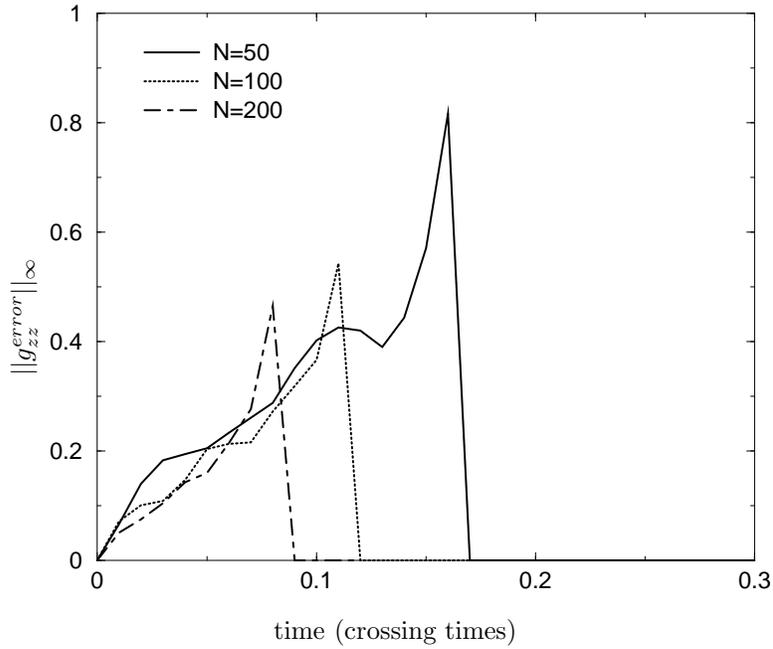}
\end{psfrags}
}
\hspace{0.1in}
\caption{The nonlinear gauge wave with shift test for the weakly hyperbolic
harmonic system. The code crashes in less than a crossing time. }
\label{fig:gwshH.cmp}
\end{figure}

\begin{figure}
\centerline{
\begin{psfrags}
\psfrag{xlabel}[ct]{time (crossing times)}
\psfrag{ylabel}[cb]{$||g_{zz}^{error}||_{\infty}$}
\epsfxsize=4in\epsfbox{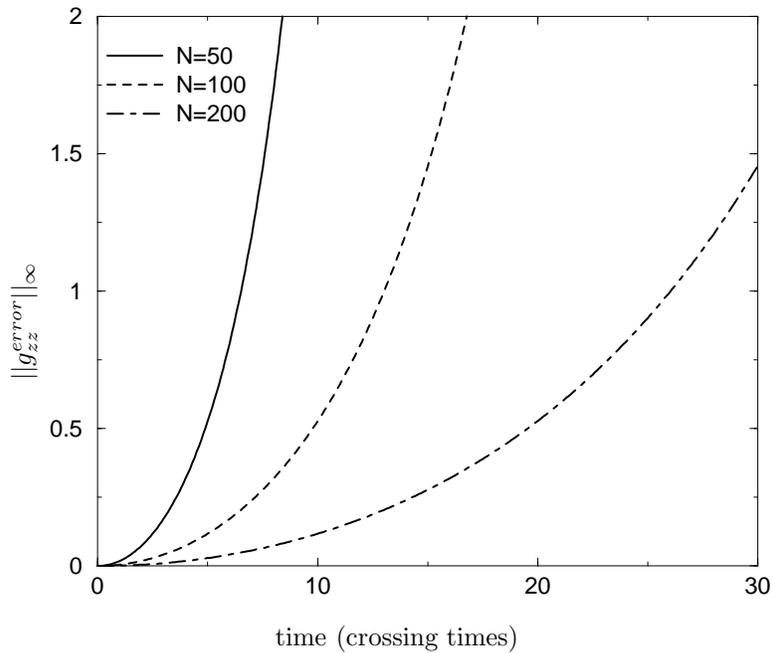}
\end{psfrags}
}
\hspace{0.1in}
\caption{The nonlinear gauge wave without shift test for the weakly hyperbolic
harmonic system, run in accord with the Apples With Apples specifications. The
convergence of the error is deceptive of code reliability. }
\label{fig:gw1DH.cmp}
\end{figure}

\begin{figure}
\centerline{
\begin{psfrags}
\psfrag{xlabel}[ct]{time (crossing times)}
\psfrag{ylabel}[cb]{$||ham||_{\infty}$}
\epsfxsize=4in\epsfbox{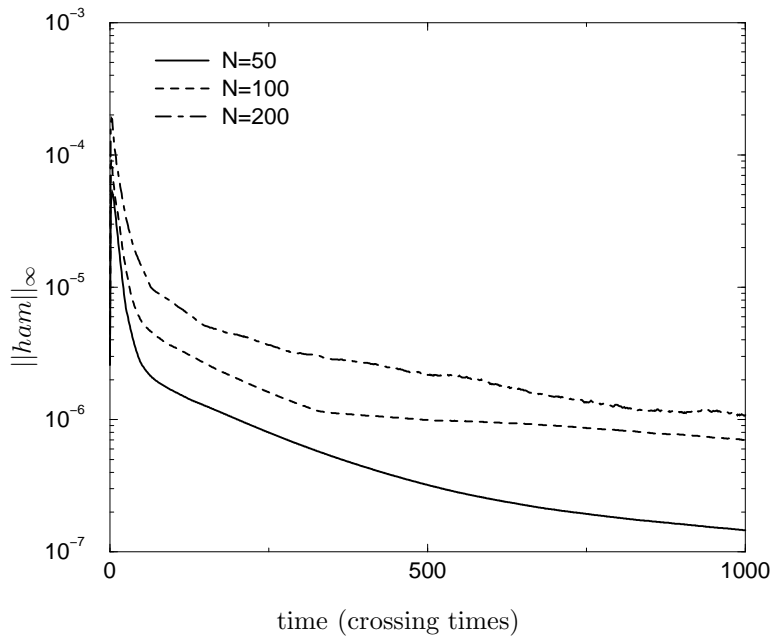}
\end{psfrags}
}
\hspace{0.1in} \caption{The robust stability test for the weakly hyperbolic
harmonic system with dissipation. As in Fig. \ref{fig:robustH.cmp}, the
Hamiltonian constraint is plotted on a linear-logarithmic scale. The
dissipation now kills the exponential growth but the growth of constraint
violation with resolution indicates that the code fails the test. } 
\label{fig:robustdH.cmp}
\end{figure}

It is instructive to investigate the performance of a code based upon this
weakly hyperbolic harmonic system by using the Apples with Apples testbed.
Figure \ref{fig:robustH.cmp} shows the results of the robust stability test,
where a simulation in the linear regime is carried out with random (constraint
violating) initial data. The results show an exponential rise in the violation
of the Hamiltonian constraint {\em  at a rate that increases with grid
resolution}, which eventually leads to a code crash. This behavior is
symptomatic of weakly hyperbolic systems and presages possible problems in the
nonlinear domain. The simulation of a nonlinear gauge wave with shift, shown in
Fig. \ref{fig:gwshH.cmp}, verifies such problems. These problems do not appear
for the nonlinear gauge wave without shift, as the results shown in Fig.
\ref{fig:gw1DH.cmp} indicate convergence. Also, as illustrated in Fig.
\ref{fig:robustdH.cmp}, with the addition of numerical dissipation, the
Hamiltonian constraint no longer grows exponentially in the robust stability
test, although the constraint violation still increases with grid resolution,
indicating failure of the test. Similar conclusions follow from the Apples with
Apples Gowdy wave tests.

These results show that a full battery of tests are necessary in order to
establish reliable code performance. Otherwise, misleading information about
code performance can result. As history has shown in the case of ADM evolution
codes, weakly hyperbolic systems system cannot be expected to give reliable
long term performance in the presence of strong fields, which makes them
unsuitable for black hole simulations. 

The Friedrich-Nagy system and the weakly hyperbolic harmonic system represent
two extremes of a dilemma facing code development in numerical relativity. On
one hand, the Friedrich-Nagy system has all the desired analytic features but
its complexity poses a barrier to code development. On the other
hand, the weakly hyperbolic harmonic system is simple and easily implemented as
an efficient code, but well-posedness is questionable. Should you try to fix
these simple systems or should you bite the bullet and develop codes based upon
formulations where a well-posed nonlinear IBVP has been fully established?  To
date, the effort in numerical relativity has been weighted heavily toward the
simpler formulations. It is timely that some attention be given to
investigating whether the Friedrich-Nagy system can be converted into a
workable code. A useful starting point would be a linearized version evolving
on $T^2\times R$, where the complications of the equations and the boundary
gauge would greatly simplify and would perhaps lead to a better understanding
of the essential elements of the approach. Most of the effort in the field can
be expected to remain a compromise between these extremes, e.g. the strongly
hyperbolic harmonic system for which a Sommerfeld boundary condition is not
constraint preserving. In all such endeavors, a close working combination of
analytic and numerical insight can offer valuable guidance.

\begin{acknowledgments}

Much of the material in this paper originated from interactions with the groups
carrying out the Apples with Apples standardized tests. We are particularly
grateful to L. Lehner, C. Palenzuela and M. Tiglio  for sharing ideas and
results. These interactions demonstrate how code comparison can  can be very
effective in tackling numerical problems. We thank B. Schmidt for several useful
comments on the manuscript. The computer simulations were carried out using
Cactus infrastructure. The hospitality of the Heraeus foundation at the
Physikzentrum Bad Honnef during part of this work was greatly appreciated. The
research was supported by National Science Foundation Grant PHY-0244673 to the
University of Pittsburgh.

\end{acknowledgments}

\end{document}